\documentclass[showpacs,preprintnumbers,amsmath,amssymb]{revtex4}
\usepackage{graphicx}% Include figure files
\usepackage{dcolumn}% Align table columns on decimal point
\usepackage{bm}% bold math
\usepackage{amsmath}
\usepackage{multirow}
\usepackage{tikz}

\def\xa{{x_\alpha}}
\def\xb{{x_\beta}}
\def\xc{{x_\gamma}}
\def\xd{{x_\delta}}
\def\yac{{y_{\alpha\gamma}}}

\def\vab{{v_{\alpha\beta}}}
\def\uacd{{u_{\alpha\gamma\delta}}}
\def\zabcd{{z_{\alpha\beta\gamma\delta}}}
\def\kB{{k_{\rm B}}}
\def\Point{{\rm Point}}

\def\ac{{\alpha\gamma}}
\def\abcd{{\alpha\beta\gamma\delta}}
\def\eg{{\em e.g. }}

\begin{document}

\title{Cluster variation method analysis of correlations and entropy in BCC solid solutions}
\author{Nathaniel Hoffman}
\author{Michael Widom}
\affiliation{Department of Physics, Carnegie Mellon University, Pittsburgh PA 15213}

\date{\today}

\begin{abstract}
Solid solutions occur when multiple chemical species share sites of a common crystal lattice. Although the single site occupation is random, chemical interaction preferences bias the occupation probabilities of neighboring sites, and this bias reduced the entropy of mixing below its ideal value. Sufficiently strong bias leads to symmetry-breaking phase transitions. We apply the cluster variation method to explore solid solutions on body centered cubic lattices in the context of two specific compounds that exhibit opposite ordering trends. Employing density functional theory to model the energetics, we show that CuZn exhibits an order-disorder transition to the CsCl prototype structure, while AlLi instead takes the NaTl prototype structure, and we evaluate their temperature-dependent order parameters, correlations and entropies.
\end{abstract}

\maketitle

\section{Introduction}

\begin{figure}
  \includegraphics[width=6in]{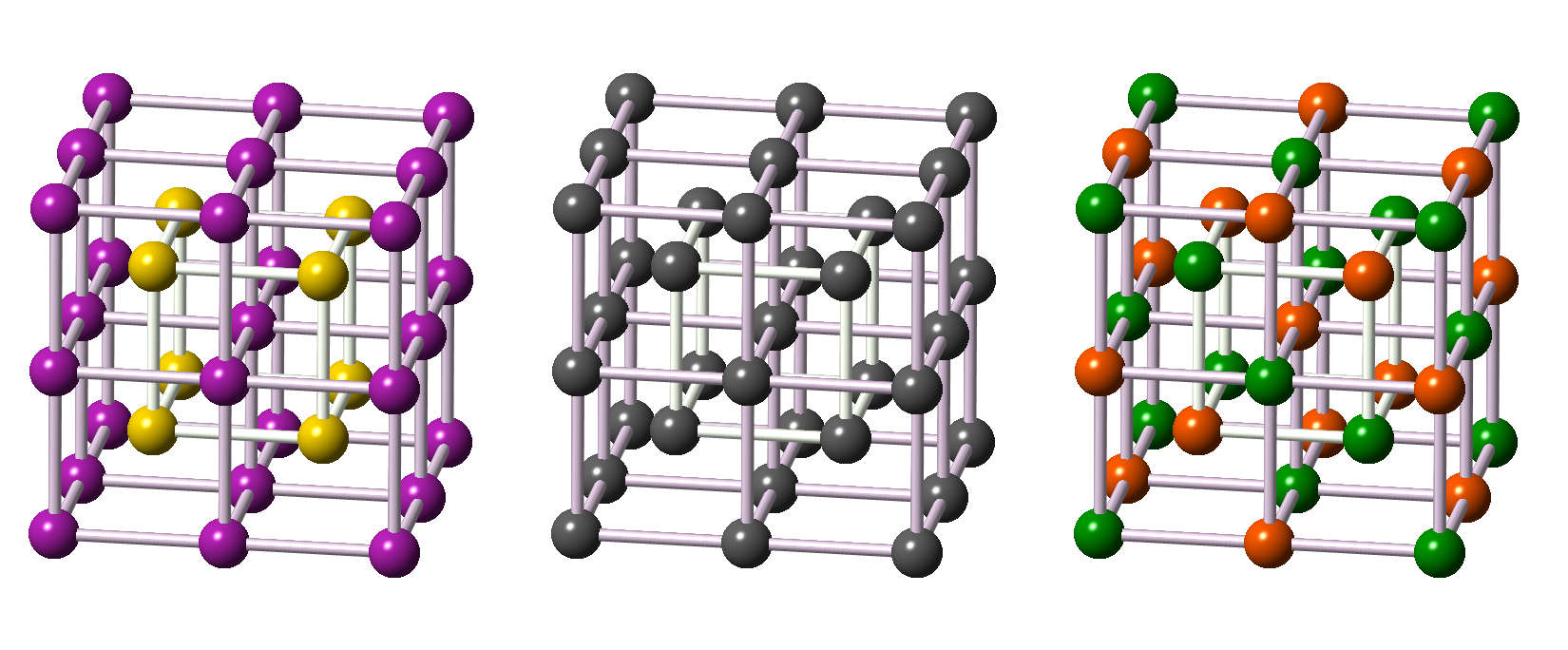}
  \caption{\label{fig:cF16}Examples of chemical order on a body centered cubic lattice. The CsCl prototype (left) with Pearson type cP2 and Strukturbericht symbol B2, and the NaTl prototype (right) with Pearson type cF16 and Strukturbericht symbol B32a, break the symmetry of the body centered cubic W prototype (center) with Pearson type cI2 and Strukturbericht symbol A2.}
\end{figure}

The body centered cubic structure (BCC, also known by its Pearson type cI2 and its Strukturbericht symbol A2) forms a crystal lattice in which all sites are equivalent. Many pure metallic elements take this structure, for example W, the crystallographic prototype. Additionally, many metal alloys adopt this structure at high temperatures, for example CuZn. In such a case each site is randomly occupied by either Cu or Zn, so that the average occupations of each site remain equivalent and the structure remains BCC from a symmetry perspective. However, each element has its own chemical interaction preferences, so that the instantaneous occupation of one site biases the occupation probability of its neighboring sites. In the case of CuZn, the preference is for unlike neighbors; at low temperatures CuZn undergoes a continuous phase transition to a chemically ordered structure of prototype CsCl (Pearson type cP2, Strukturbericht B2). Other compounds exhibit differing preferences. For example, AlLi favors a pattern of like nearest neighbors but unlike second neighbors and acquires the ordered structure of prototype NaTl (Pearson type cF16, Strukturbericht B32a). Fig.~\ref{fig:cF16} illustrates these three structures.

While local chemical order can reduce the internal energy $U$, the resulting biased occupation probabilities reduce the entropy $S$. The trade-off between energy and entropy is captured by the free energy $F=U-TS$. While energies are readily calculated using electronic density functional theory (DFT), entropy calculations require supplementing the DFT-based energies with methods of statistical mechanics. Here we apply a sequence of approximations related to Kikuchi's Cluster Variation Method~\cite{Kikuchi1951,deFontaine79,TurchiCuZn,deFontaine94} in order to minimize the free energy and thereby calculate the temperature-dependent correlation functions, order parameters and entropies. We derive our entropy expressions heuristically based on the information content of correlation functions and the correspondence between information and entropy. Specifically, we utilize a single point approximation, a pairwise correction, and a four-point approximation~\cite{Ackermann89}. Although the methods presented here are applied only to binary alloys, they readily generalize to more complex alloy systems such as Heusler compounds~\cite{FelserBook2016} and high entropy alloys~\cite{Yeh04_1,Cantor04}.

\section{Methods}

\subsection{Correlation functions}
\label{sec:Correlations}

\begin{figure}
\includegraphics[width=3in]{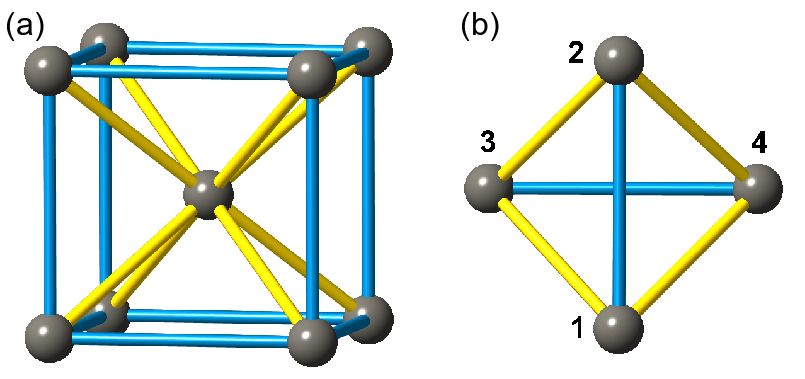}
\caption{\label{fig:BCC}(a) The body centered cubic unit cell contains two equivalent sites (cube vertex and center). Nearest neighbor bonds are shown in yellow and next nearest neighbors in blue. (b) The BCC tetrahedron with specific sites labeled $\alpha-\delta$.}
\end{figure}

Let Greek indices label chemical species, and $\xa$ be the mole fraction of species $\alpha$ in the solid solution. In the absence of other information, $\xa$ is the probability that a given lattice site is occupied by species $\alpha$. As shown in Fig.~\ref{fig:BCC}a, each site of the BCC lattice has 8 nearest neighbors (\eg the bond $1-3$ in Fig.~\ref{fig:BCC}b). Denote by $\yac$ the fraction of nearest neighbor bonds that have species $\alpha$ on one end and $\gamma$ on the other. Each site has 6 next-nearest neighbors at a distance only $2/\sqrt{3}=1.15\times$ the nearest neighbor separation (\eg the bond $1-2$), joining species $\alpha$ and $\beta$ with probability $\vab$. Also present in the BCC structure are isosceles triangles with two nearest and one next-nearest neighbor bond as edges (\eg the triangle $1-3-4$). We denote the probability of species $\alpha$ on the symmetric vertex ($1$) and species $\gamma$ and $\delta$ on the others ($3$ and $4$) as $\uacd$. Finally, we define the four point function $\zabcd$ as the probability for species $\alpha$ on site $1$, $\beta$ on $2$, $\gamma$ on $3$ and $\delta$ on $4$.

Notice the sum-rule relationships among the probabilities. For example, $\uacd=\sum_\beta \zabcd$, $\yac=\sum_\delta \uacd$, $\xa=\sum_\gamma\yac$, and finally $\sum_\alpha\xa=1$ expresses the normalization. In the limit of perfect disorder, when each site is occupied independently of its neighbors, the probabilities factorize, \eg $\yac=\xa\xc$, and $\zabcd=\xa\xb\xc\xd$. In general the site occupations are {\em not} independent and we refer to the nontrivial probabilities as {\em correlation functions}.

\subsection{Interaction model}
We may express the energy in terms of the correlation functions defined above. The most general expression for the energy per atom is
\begin{equation}
  \label{eq:H}
  U=6\sum_\abcd E_\abcd ~z_\abcd,
\end{equation}
where the factor of 6 arises because the BCC lattice has 6 tetrahedra/atom. Linear combinations of the coefficients $E_\abcd$ correspond to the energies of individual chemical species, pairwise, and three- and four-body interactions. As noted above, the next-nearest neighbor bond is only 15\% longer than the nearest neighbor, while the third neighbor is twice as long. It is reasonable to keep nearest and next nearest neighbor interactions, but to omit further neighbors. Since both alloy systems are well described as nearly free electron systems, it is reasonable to neglect many-body interactions as a first approximation.

For a binary alloy at fixed composition, the fraction of nearest neighbor bonds joining like species is linearly dependent on the number joining unlike species, and the same is true for next-neighbor bonds. Thus it suffices to parameterize the energy in terms of parameters $J$ and $K$ representing the energies of unlike nearest and next-nearest neighbor bonds. Formally, note that
\begin{align}
  \sum_\ac E_\ac ~\yac &= \sum_{\alpha\ne\gamma}E_\ac ~\yac
                + \sum_\alpha E_{\alpha\alpha} ~y_{\alpha\alpha} \\ \nonumber
                      &= \sum_{\alpha\ne\gamma}\left(E_\ac-\frac{1}{2}(E_{\alpha\alpha}+E_{\gamma\gamma})\right) ~\yac
                        + \frac{1}{2}\left(\sum_\alpha E_{\alpha\alpha} ~\xa + \sum_\gamma E_{\gamma\gamma} ~\xc\right).
\end{align}
Only off-diagonal terms ($\alpha\ne\gamma$) multiply $\yac$, and the final term in the above depends only on global composition and hence is irrelevant. We define $J=E_\ac-(E_{\alpha\alpha}+E_{\gamma\gamma})/2$ for nearest neighbor bonds, and similarly define $K$ for next-nearest neighbor bonds. Finally, counting the numbers of unlike bonds in the set of decorated tetrahedra, we find that
\begin{equation}
  \label{eq:Eabcd}
  E_\abcd=\frac{1}{6}J(4-\delta_{\alpha\gamma}-\delta_{\alpha\delta}-\delta_{\beta\gamma}-\delta_{\beta\delta})
  +\frac{1}{4}K(2-\delta_{\alpha\beta}-\delta_{\gamma\delta}).
\end{equation}

To determine the interaction parameters $J$ and $K$, we decorate the BCC lattice with different atomic species in various configurations and calculate the energies using electronic density functional theory (DFT). We then fit the results to Eq.~(\ref{eq:Eabcd}). Our DFT calculations are performed in the PBE generalized gradient approximation~\cite{Perdew96} using projector augmented wave potentials~\cite{Kresse99} as implemented in VASP~\cite{VASP}. Default plane wave energy cutoffs are applied and we choose $k$-point meshes to achieve energy convergence to better than 1 meV/atom. As our ensemble of structures we take the set of all symmetry-inequivalent equiatomic decorations of a 16-atom cF16 unit cell, based on the experimental lattice parameters. We perform two sets of calculations, one holding the atoms at ideal BCC lattice positions, and the other employing full relaxation of lattice parameters and atomic coordinates.

Given our complete set of decorated cF16 unit cells, we may determine the energy minimizing structure within this set as functions of the interaction parameters $J$ and $K$. Note that only the signs and ratio of  $J$ and $K$ matter. We find that two structures minimize the energy across the majority of the  $J$ and $K$ plane. Negative $J$ tends to favor unlike near neighbors, as occurs in the CsCl prototype, while negative $K$ favors unlike next-neighbors as occurs in the NaTl prototype. Because of the relative numbers of nearest and next-nearest neighbors, the crossover occurs at $J=3K/2$. The situation for positive J and K is more complex, and we find a variety of non-cubic structures minimize the energy for $K>0$ and $J>K/2$.

\begin{figure}
\includegraphics[width=3in]{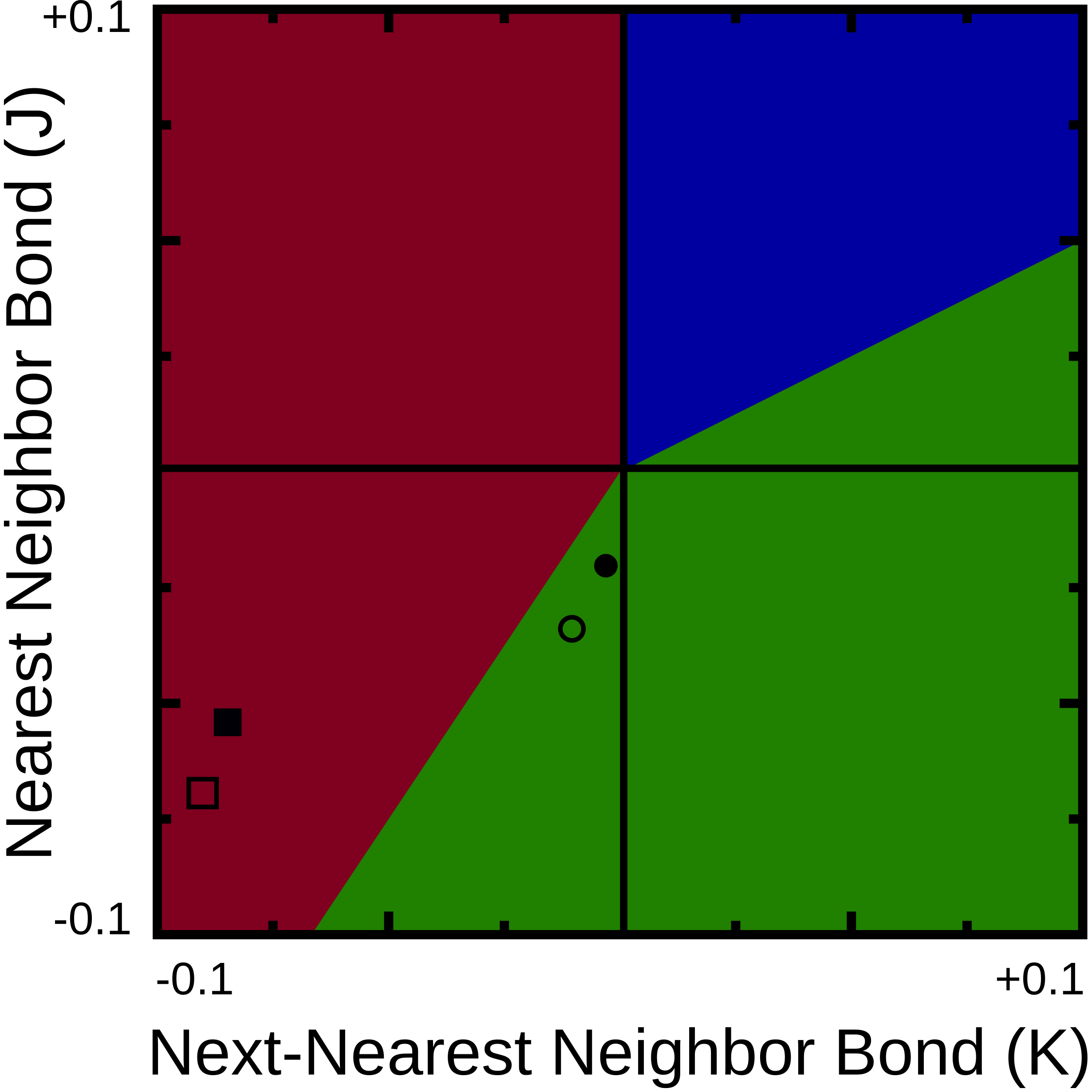}
\caption{\label{fig:JK}Energy minimizing structure as functions of $J$ and $K$ (units eV/bond). Green is NaTl, red is CsCl, and blue contains various non-cubic decorations. Fitted values for CuZn (circles) and AlLi (squares) are marked. Filled symbols are relaxed, open symbols are unrelaxed.}
\end{figure}

\begin{figure}
\includegraphics[width=5in]{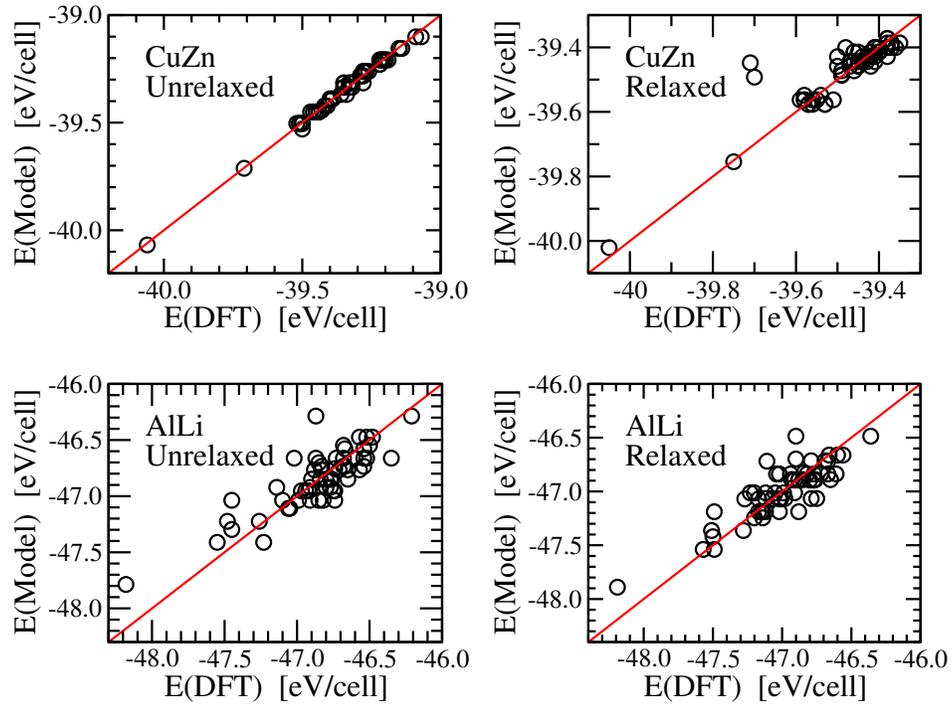}
\caption{\label{fig:fits}Fits to the JK model. The two outliers in relaxed CuZn are excluded because their structures underwent strong shear deformations.}
\end{figure}

Fits to the JK model are illustrated in Fig.~\ref{fig:fits}, and resulting values of $J$ and $K$ are given in Table~\ref{tab:fits}. In the case of CuZn, the nearest neighbor interaction dominates and favors alternation of species as in the CsCl prototype. AlLi, in contrast, is strongly ionic. Owing to the Coulomb interaction, the next-nearest neighbor interaction is comparable in strength to the nearest neighbor and also favors alternation of species, as in the NaTl prototype.

\begin{table}
  \caption{\label{tab:fits}Interaction parameters for CuZn and AlLi as calculated within DFT. Units are eV/bond.}
  \begin{tabular}{r|ll|ll}
        & \multicolumn{2}{c|}{Unrelaxed}&\multicolumn{2}{c}{Relaxed}\\
    \hline
    CuZn& J = -0.0369, & K = -0.0134 & J = -0.0228, & K = -0.0056 \\
    AlLi& J = -0.0729, & K = -0.0937 & J = -0.0571, & K = -0.0878 \\
  \end{tabular}
\end{table}

\subsection{Cluster variation model}

Kikuchi's Cluster Variation Method (CVM)~\cite{Kikuchi1951,deFontaine79,Ackermann89,Widom16} supplements the interaction energy model with a model for the entropy. We present a heuristic derivation of the entropy based on its equivalence (in suitable units) to information. Consider a crystalline solid containing multiple chemical species, with each species $\alpha$ present in mole fraction $\xa$. If the species are distributed uniformly across $N$ sites then we must specify the $\xa N$ sites occupied by each species $\alpha$. Equivalently we must specify precisely one of the $\Omega_P = N!/\prod_\alpha (\xa N)!$ distinct configurations. Imagine an index listing every possible configuration. The number of digits to specify an index entry is the logarithm of $\Omega_P$. Applying the Stirling expansion yields the ideal (Bragg-Williams~\cite{Bragg34}) entropy per site \begin{equation}
\label{eq:SPoint}
S_{\rm Point}=-\sum_\alpha \xa\ln{\xa}
\end{equation} 
where we have chosen to measure the entropy in units of the Boltzmann constant $\kB$.  

Correlations reduce the entropy from its ideal value because knowledge of the chemical species on a given site provides information about the likely occupation of nearby sites. In view of the equivalence of information and entropy, if we can quantify this information we can then subtract it from $S_\Point$. This approach is nontrivial because we must determine the {\em minimum} information contained in the configuration, avoiding the inclusion of redundant information. If the distribution of atomic species is correlated, for example if the probability $\yac$ to find species $\alpha$ and $\gamma$ as nearest neighbors differs from the uncorrelated expectation of $\xa\xc$, then the {\em mutual information} per bond,
\begin{equation}
\label{eq:mutual}
I[\{\yac\}] =\sum_\ac \yac \ln{\yac/\xa\xb},
\end{equation}
must be subtracted from the ideal entropy yielding
\begin{equation}
\label{eq:S-I}
S_{\rm Pair}=S_\Point[\{\xa\}]-\frac{z}{2} I[\{\yac\}],
\end{equation}
where $z$ is the coordination number of the lattice. Specifying the correlation $\yac$ reduces the additional information required to specify a full configuration. In thermodynamics Eq.~(\ref{eq:S-I}) is known as the Bethe-Guggenheim quasichemical approximation~\cite{Bethe35,Guggenheim44}. For the BCC lattice, $z=8$, and the entropy becomes
\begin{equation}
  \label{eq:Spair}
  S_{\rm Pair} = 7\sum_i \xa\ln\xa-4\sum_\ac\yac\ln\yac
\end{equation}
at the level of nearest neighbor pair correlations.

The pair correlation function $\yac$ is equivalent to the Warren-Cowley short-range order parameter $\alpha_\ac\equiv 1-\yac/\xa\xc$. In view of the pair-level approximation Eq.~\ref{eq:S-I}, we may express the entropy in terms of concentrations and the Warren-Cowley parameter as
\begin{equation}
  \label{eq:WC}
  S_{\rm WC}=-\sum_\alpha \xa\ln{\xa}
  -\frac{z}{2}\sum_\ac \xa\xc(1-\alpha_\ac)\ln{(1-\alpha_\ac)}.
\end{equation}
Such an approximation could be applied regardless of how the order parameter was obtained, including experimental determination.

\def\CVMone{{\{1\}}}

\def\CVMdot{{\tikz{\filldraw (0,0) circle (0.05cm);}}}

\def\CVMnn{{
\tikz[baseline]{
\draw (0,-.05)--(.2,.15);
\filldraw (0,-.05) circle (0.05cm);
\filldraw (.2,.15) circle (0.05cm);}}}

\def\CVMnnn{{
\tikz{\draw (0,0)--(.4,0);
\filldraw (0,0) circle (0.05cm);
\filldraw (.4,0) circle (0.05cm);}}}

\def\CVMtri{{
\tikz[baseline]{
\draw (0,-.05)--(.2,.15);
\draw (0,-.05)--(.4,-.05);
\draw (.4,-.05)--(.2,.15);
\filldraw (0,-.05) circle (0.05cm);
\filldraw (.2,.15) circle (0.05cm);
\filldraw (.4,-.05) circle (0.05cm);}}}

\def\CVMtet{{
\tikz[baseline]{
\draw (0,.05)--(.2,.25);
\draw (0,.05)--(.2,-.15);
\draw (.4,.05)--(.2,.25);
\draw (.4,.05)--(.2,-.15);
\draw (.2,-.15)--(.2,.25);
\draw (0,0.05)--(0.15,0.05);
\draw (0.25,0.05)--(0.4,0.05);
\filldraw (0,0.05) circle (0.05cm);
\filldraw (.2,.25) circle (0.05cm);
\filldraw (.2,-.15) circle (0.05cm);
\filldraw (.4,.05) circle (0.05cm);}}}

The mutual information approach requires further refinement because additional correlations may be present beyond nearest-neighbor or among multiple neighbors. Even the bond probabilities $\yac$ may contain redundant information. Eqs.~(\ref{eq:SPoint}) and~(\ref{eq:S-I}) represent the first steps in a systematic expansion of the entropy involving successively longer-range and higher-order correlation functions~\cite{Yedidia2005,Pelizzola2005}. CVM approximates the entropy per site as $S =\frac{1}{N}\ln{\Omega}$ where $\Omega$ is given by a product of combinatorial factors reflecting the correlation functions of pairs of sites (bonds), triplets, etc. Each factor represents the information content of a correlation function relative to the information contained in its constituent parts. We define numerical values for symbols associated with the empty lattice, isolated points, nearest neighbor bonds, next-nearest neighbor bonds, triangles, and tetrahedra
\begin{equation}
\begin{split}
\label{eq:CVMdef}
\CVMone \equiv N!,~~
\{\CVMdot\}\equiv\prod_{\alpha} (x_{\alpha} N)!,~~
\{\CVMnn\}\equiv\prod_{\alpha\gamma} (y_{\alpha\gamma} N)!,\\
\{\CVMnnn\}\equiv\prod_{\alpha\beta} (v_{\alpha\beta} N)!,~~
\{\CVMtri\}\equiv\prod_{\alpha\gamma\delta} (u_{\alpha\gamma\delta} N)!,~~
\{\CVMtet\}\equiv\prod_{\alpha\beta\gamma\delta} (z_{\alpha\beta\gamma\delta} N)!.
\end{split}
\end{equation}

Defining the combinatorial factors
\begin{equation}
\setlength{\jot}{10pt} % affecting the line spacing in the environment
\begin{split}
\label{CVMW}
f(\CVMdot)=\frac{\CVMone}{\{\CVMdot\}},~~~
f(\CVMnn)=\frac{\{\CVMdot\}^2}{\{\CVMnn\}\CVMone},~~~
f(\CVMnnn)=\frac{\{\CVMdot\}^{2}}{\{\CVMnnn\}\CVMone}, \\
f(\CVMtri)=\frac{\{\CVMnn\}^2\{\CVMnnn\}\CVMone}{\{\CVMdot\}^3\{\CVMtri\}},~~~
f(\CVMtet)=\frac{\{\CVMdot\}^4\{\CVMtri\}^{4}\CVMone}{\{\CVMnn\}^4\{\CVMnnn\}^2\{\CVMtet\}},
\end{split}
\end{equation}
we recognize the ``point'' approximation $\Omega_P=f(\CVMdot)$, as the Bragg-Williams~\cite{Bragg34} formula, while the nearest neighbor pairwise correction $\Omega_{\rm NN}=f(\CVMdot)f^m(\CVMnn)$ is the Bethe-Guggenheim result ($m=4$ for BCC). Next-nearest neighbors multiply $\Omega$ by $f^3(\CVMnnn)$ and triangles provide an extra factor of $f^{12}(\CVMtri)$. Finally, including all factors, we obtain
\begin{equation}
\begin{aligned}
\Omega_{\rm TET}
&= f(\CVMdot) f^4(\CVMnn) f^3(\CVMnnn) f^{12}(\CVMtri) f^6(\CVMtet) \\
&= \frac{\CVMdot^1\CVMtri^{12}}{\CVMnn^4\CVMnnn^3\CVMtet^6}
\end{aligned}
\end{equation}
which yields the entropy approximation for BCC lattices
\begin{equation}
\begin{aligned}
\label{eq:S_tet}
S &= \sum_\alpha \xa\ln{\xa}
    +12\sum_{\alpha\gamma\delta}u_{\alpha\gamma\delta}\ln{u_{\alpha\gamma\delta}}\\
    &-4\sum_{\alpha\beta}\yac\ln{\yac}
    -3\sum_{\alpha\beta}v_{\alpha\beta}\ln{v_{\alpha\beta}}
    -6\sum_{\alpha\beta\gamma\delta} z_{\alpha\beta\gamma\delta}\ln{z_{\alpha\beta\gamma\delta}}.
  \end{aligned}
  \end{equation}
This expression is equivalent to that previously presented by Kikuchi~\cite{Kikuchi1987,Ackermann89} for BCC structures.

We now have a sequence of higher-order approximations to the entropy, starting with the point approximation Eq.~(\ref{eq:SPoint}), through the pair approximation Eq.~(\ref{eq:Spair}), to the tetrahedron approximation Eq.~(\ref{eq:S_tet}). Note that we recover the lower approximations from the higher by making {\em superposition} approximations. Specifically, the substitutions $\zabcd=\yac y_{\beta\delta}$, $\uacd=\yac \xd$, and $\vab=\xa\xb$ reduce the tetrahedron approximation~(\ref{eq:S_tet}) to the pair approximation~(\ref{eq:Spair}). Similarly, the substitution $\yac=\xa\xc$ reduces the pair approximation to the point approximation (\ref{eq:SPoint}).

\section{Results}

When combined with the energy Eq.~(\ref{eq:H}) we obtain the free energy $F=E-TS$. We now minimize the free energy with respect to the correlation functions in order to predict their temperature-dependent values. Differentiating $F$ with respect to the highest order correlation function (and imposing the sum rules discussed in Section~\ref{sec:Correlations}) yields an identity for this function in terms of lower order functions. Summing the higher-order function to produce new lower-order functions, and iterating this procedure, results in convergence towards a self-consistent solution for all correlation functions~\cite{Kikuchi1974}.

For the specific case of BCC, we reproduce the result from Ref.~\cite{Ackermann89} (with some slight notation changes)
\begin{equation}
  \label{eq:iter}
  \zabcd = e^{-\lambda/6\kB T}e^{-E_\abcd/\kB T}X^{1/24}U^{1/2}Y^{-1/6}V^{-1/4}
\end{equation}
with
\begin{equation}
  \label{eq:products}
  \begin{aligned}
    X &= \xa\xb\xc\xd &
    U &= \uacd u_{\beta\gamma\delta}u_{\alpha\beta\gamma}u_{\alpha\beta\delta}\\
    Y &= \yac y_{\alpha\delta} y_{\beta\gamma} y_{\beta\delta} &
    V &= \vab v_{\gamma\delta}.
  \end{aligned}
\end{equation}
In each correlation function above, the species index $\alpha-\delta$ applies, respectively, to the tetrahedron vertices $1-4$ as shown in Fig.~\ref{fig:BCC}. Note that by breaking the equivalence of lattice sites we allow for the possibility of spontaneous symmetry breaking. We recognize the exponents in Eq.~(\ref{eq:iter}) as the coefficients in the entropy Eq.~(\ref{eq:S_tet}) divided by 6 times the number of correlation factors in Eq.~(\ref{eq:products}). The constant $\lambda$ is related to the Lagrange multiplier and is determined by the required normalization of $\zabcd$.

\begin{figure}
  \includegraphics[width=3in]{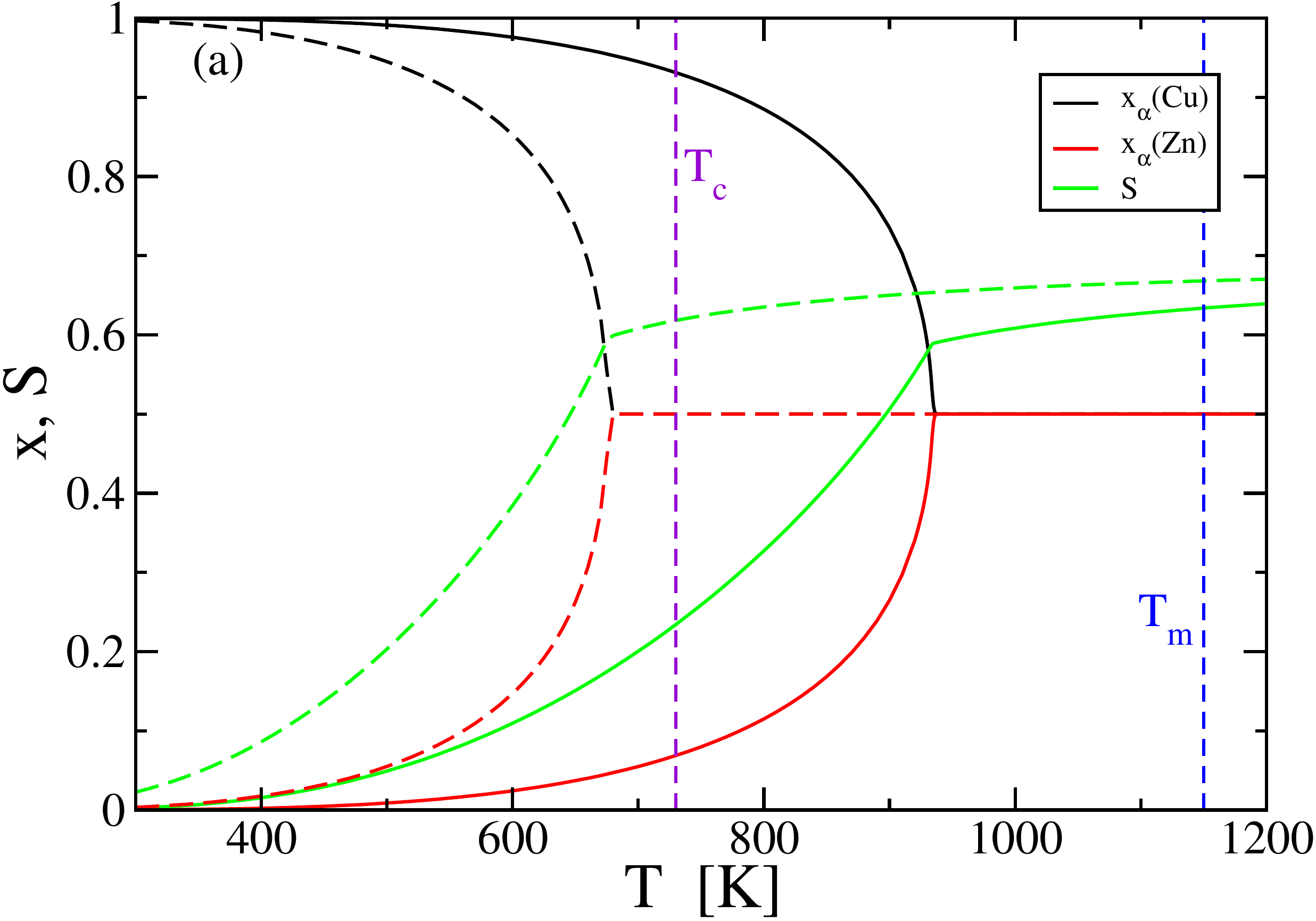}
  \includegraphics[width=3in]{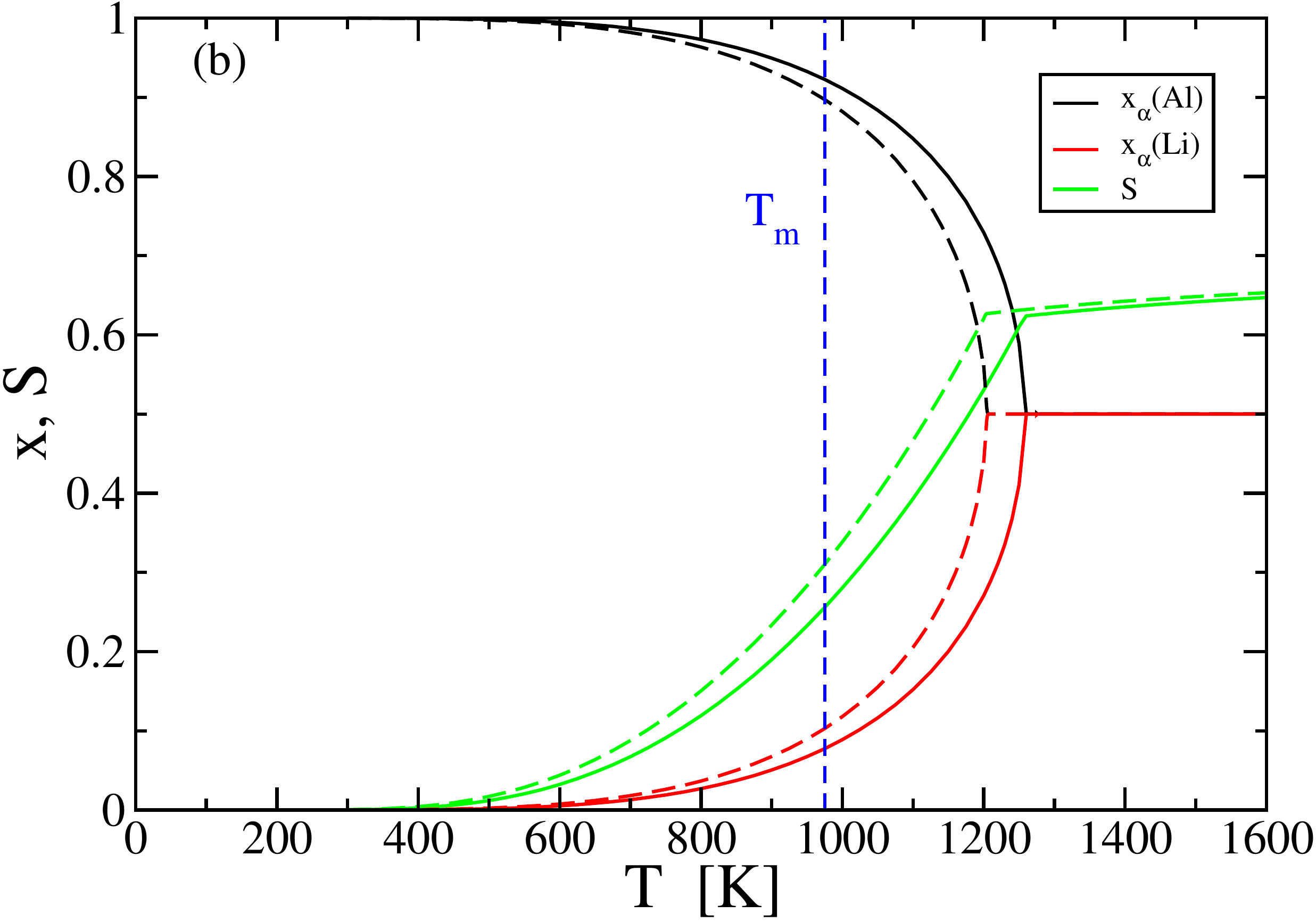}
  \caption{\label{fig:XofT}(a) Order parameters $\xa(\rm Cu)$ and $\xc(\rm Zn)$ and entropy $S$ (in units of $\kB$) of CuZn obtained from the tetrahedron method utilizing fitted values of the unrelaxed (solid lines) and relaxed (dashed lines) interactions as given in Table~\ref{tab:fits}. (b) $\xa(\rm Al)$, $\sb(\rm Li)$, and $S$, for AlLi.}
\end{figure}

Fig.~\ref{fig:XofT}a shows the result of the tetrahedron approximation for CuZn. We plot the point order parameters $x_{\rm Cu}$ at the vertex and body center sites. Above a critical temperature $T_c$ both are identically 1/2, while below $T_c$ the symmetry spontaneously breaks, yielding the CsCl structure. We also plot the entropy, which grows monotonically from zero at low $T$ towards its ideal maximum value $\ln{2}$ at high $T$. We carry out the calculation first using unrelaxed parameters, resulting in $T_c=936$K, and then again using relaxed parameters, resulting in $T_c=681$K. The experimental $T_c\sim 730$K. The discrepancy between our relaxed CVM $T_c$ and the experimental value is likely due to neglected effects of thermal expansion, and vibrational~\cite{TurchiCuZn} and electronic free energies.

Fig.~\ref{fig:XofT}b displays the corresponding result for AlLi. Here, $3K/2<J<0$, so the system takes the NaTl prototype structure, so we plot $x({\rm Al})$ at positions $a$ and $b$. In this case, the predicted order-disorder transition lies somewhat above the experimental melting point, implying that chemical order persists for all temperatures below melting, consistent with experimental observation. 

\begin{figure}
  \includegraphics[width=3in]{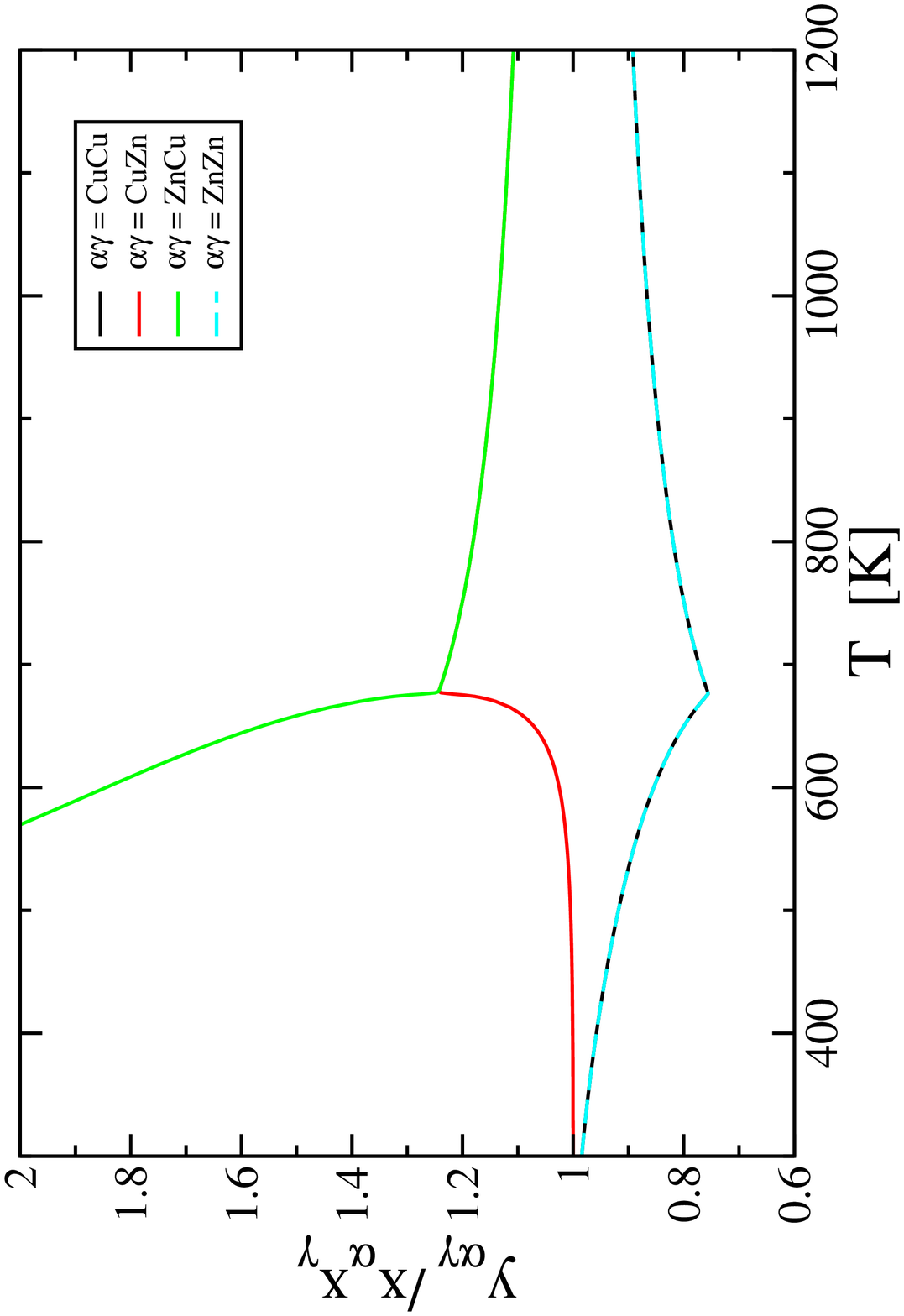}
  \caption{\label{fig:YoX}Ratios $\yac/\xa\xc$ for CuZn using relaxed fit parameters.}
\end{figure}

Note that entropy remains below $\ln{2}$ even in the disordered symmetric state above $T_c$. This occurs because of the pair and multi-point correlations, which differ from products of single-point functions, as illustrated in Fig.~\ref{fig:YoX} where we plot the ratio $\yac/\xa\xc$. Note that all four elements of $\yac$ remain nonzero at all $T$. They slowly approach the uncorrelated values $\xa\xc$ at high $T$ but remain far from those values over the range plotted. Two values ($y_{\rm CuCu}$ and $y_{\rm ZnZn}$) remain below the uncorrelated value because of the unfavorable like-like interaction at nearest neighbors, while the other two ($y_{\rm CuZn}$ and $y_{\rm ZnCu}$) lie above owing to the favorable unlike interaction. As the site occupations $\xa({\rm Cu})$ and $\xc({\rm Zn})$ approach 1 (full occupation) at low $T$, $y_{\rm CuZn}$ also approaches 1. In the opposite case, although $\xa({\rm Zn})$ and $\xc({\rm Cu})$ both vanish at low $T$, the ratio $\yac({\rm ZnCu})/\xa({\rm Cu})\xc({\rm Zn})$ diverges owing to the favorable Zn-Cu interaction.

\begin{figure}
  \includegraphics[width=3in]{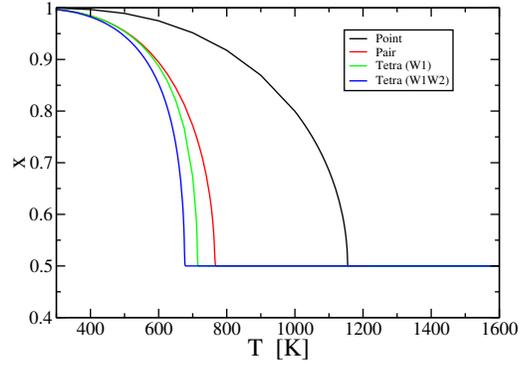}
  \caption{\label{fig:CVM-series}Series of improved approximations: point, pair and tetrahedron with a single interaction parameter$J'$, and tetrahedron with two parameters $J,K$.}
\end{figure}

Finally, we explore the relative capabilities of the series of CVM approximations, point, pair and tetrahedron. Since point and pair involve only the nearest-neighbor interaction, we re-fit the set of energies to obtain $J'=-0.0190$ (in the relaxed case) and define $E'_\ac=J'(1-\delta_\ac)$. The iteration procedure for pairs is
\begin{equation}
  \yac = e^{-2\lambda'/z\kB T} e^{-E'_\ac/\kB T}(\xa\xc)^{(z-1)/z)}
\end{equation}
and for points is
\begin{equation}
  \xa = e^{-2\lambda'/z\kB T} e^{-z E'_\ac \xc/\kB T}.
\end{equation}
As shown in Fig.~\ref{fig:CVM-series} the point approximation, which is conventional mean field theory, has the highest $T_c$, with the pair and then the tetrahedron approximations being progressively lower. This is because mean-field theory neglects all fluctuations and thus stabilizes the ordered phase up to higher temperatures. Likewise, the fall-off of the order parameters become progressively steeper. Although in every case the order parameter vanishes with mean-field critical exponent $\beta=1/2$, the pair and tetrahedron approximations include fluctuations at a localized level, reducing the transition temperature and yielding order parameter variation closer to the correct 3D Ising $\beta=0.326$. Similar effects have been noted for the simple cubic~\cite{Kikuchi1951} and square lattices~\cite{Pelizzola2005}.

Our series of improved approximations omits cases that could have been tried, such as a two pair (NN plus NNN) or a triangle approximation. These are known to be poor choices. Vul and de~Fontaine~\cite{Vul93} show that the optimal choice of a maximal cluster in a CVM approximation must be self contained in the sense that any additional point would be less closely connected to some existing vertex than the set of points already included. Clearly the NN plus NNN approximation implies the necessity to complete the triangle, while the an additional vertex bound to the triangle by NN bonds completes the tetrahedron.

\section{Conclusions}

We applied the cluster variation method to explore chemical ordering in two alloy families, CuZn and AlLi. Nearest and next-nearest neighbor interactions $J$ and $K$ were obtained from density functional theory calculations. We found that these parameters lie in a region of the $J-K$ space that favors CsCl-type ordering for CuZn, while NaTl-type ordering is favored for AlLi. CuZn exhibited an order-disorder transition, while the predicted $T_c$ for AlLi lay above its melting temperature. In each case we obtained the temperature variation of the single-point order parameter as well as a variety of higher-order correlation functions, and we plotted the entropy, which remains below its ideal value at all temperatures.

\begin{figure}
  \includegraphics[width=6in]{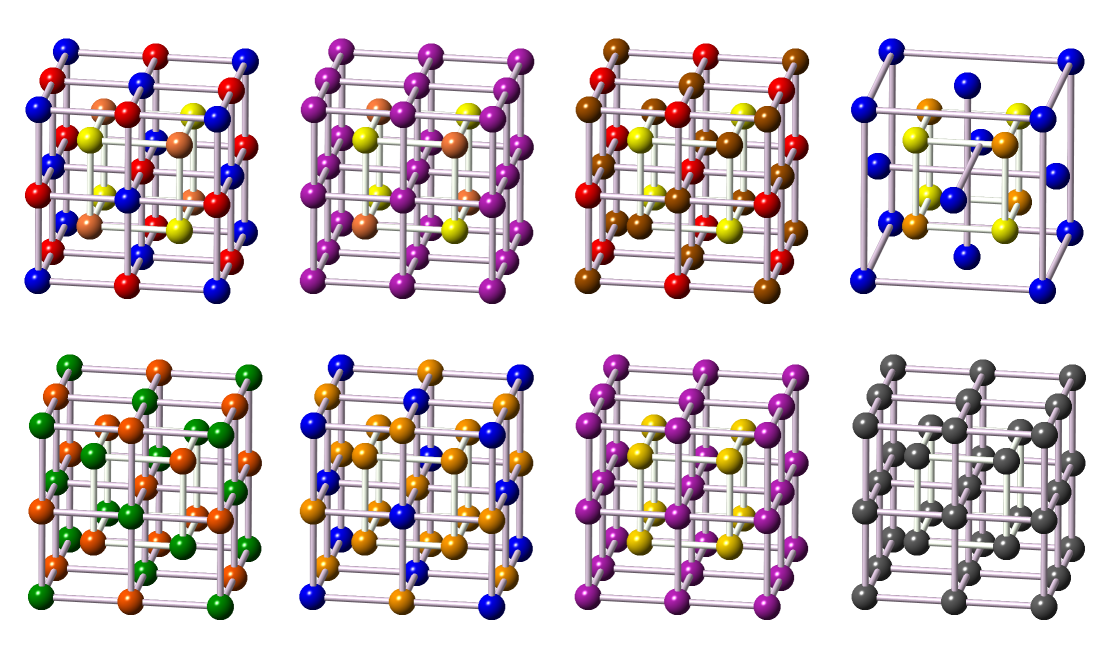}
  \caption{\label{fig:Heusler}Variants of Heusler structures. From top left to bottom right: quaternary Heusler, full Heusler, inverse Heusler, half Heusler, NaTl, BiF$_3$, CsCl, W.}
\end{figure}

Note that both the CsCl and the NaTl types of order represent different patterns of symmetry breaking starting from a disordered body centered solid solution. With additional elements, such as are present in high entropy alloys~\cite{Yeh04_1,Cantor04}, further stages of symmetry breaking are conceivable, leading to the full family of Heusler-type structures~\cite{FelserBook2016} as illustrated in Fig.~\ref{fig:Heusler}. The CVM is ideally suited to study chemical ordering in these compounds because the four interpenetrating face centered cubic sublattice of the quaternary Heusler occupy the four vertices of the BCC tetrahedron. Given an interaction model the CVM can predict ordering and mixing among the sublattices.

\section*{Acknowledgements}
This research was supported by the Department of Energy under grant DE-SC0014506.
 
%\message{here I am}

\bibliography{refs}

\end{document}